# Pressure-Constant Monte Carlo Simulation of Solid $CO_2$ Phase I up to 10 GPa using Kihara Potential Model

Koji Kobashi

Former Research Assistant, Physics Department, Colorado State University, Fort Collins, CO, USA,

and

Former Senior Researcher, Kobe Steel, Ltd., Japan

Abstract

Solid $CO_2$ phase I was studied using the pressure-constant *NPT* Monte Carlo simulation and the Kihara core potential in the temperature range of $\leq$ 194 K and the pressure range of $\leq$ 10 GPa. At a pressure of 1 bar, the temperature dependence of the calculated lattice constant agreed reasonably well with experiment. It was found that the random distribution of molecular orientations due to temperature gave a significant contribution to the increase in the lattice constant. At high pressure, the pressure dependence of the lattice constant also agreed well with experiment.

1. Introduction

This article is a continuation of the previous [1] and related articles [2-4] of the Monte Carlo (MC) simulation of solid $CO_2$ phase I using the Kihara core potential model (the Kihara potential) for intermolecular potential. Since the present article will conclude the study of solid $CO_2$ phase I using the Kihara potential and the MC simulation, it would be of use to explain the Kihara potential, which will be given in Appendix I. The algorithm of the MC simulation employed in the present work followed the one described in Ref. 5, and the computational problems encountered during the present work, which will be perhaps common to MC simulations of molecular solids, as well as the techniques used to solve them will be described in Appendix II for the reader who will engage in similar works in the future.

One of the major problems to be solved in the present work was concerned with the temperature dependence of the calculated lattice constant $a$ at the ambient pressure: in experiments [6], the observed lattice constant $a_0$ (in units of angstrom, Å, where 1 Å = 0.1 nm) was dependent on temperature $T$ (in units of Kelvin) in such a way as:

$$a_0 = 5.54 + 4.68 \times 10^{-6} T^2 \qquad (1)$$

between $T$ = 20 and 114 K. This equation roughly applies to the entire region of solid $CO_2$ phase I up to $T$ = 194 K. Preliminary computations (not shown) made after the previous article [1], however, resulted in almost linear temperature dependence of $a$, in disagreement with experiments [6]. The method used in the previous article [1] was of pragmatic use because both every molecule and the volume $V$ of the cubic basic cell that contained 2048 molecules were subject to the MC trial once in a single cycle of MC simulation: more precisely, the method consisted of a single MC trial for both the molecular center and orientation at a time for every molecule as well as a MC trial for the volume $V$. Thus, in a single cycle, 2049 MC trials were undertaken in a random order. In a single run, this cycle was repeated usually on the order of $10^8$ times. It is known, however, that this method is not rigorously correct mathematically for MC simulations [5]. Under these circumstances, the standard algorithm of the constant-*NPT* MC simulation, explained in Ref. 5, was used in the present article, and consequently an *a-T* relation like Eq. 1 was obtained, which was consistent with experiment [6]. In order to see the effect of the random distribution of molecular orientations on the *a-T* relation, MC simulations were undertaken at the ambient pressure $P = 1$ bar (1 bar = $10^{-4}$ GPa) in which molecular orientations were fixed according to the $Pa3$ structure and only molecular centers were allowed to move. It was found that the random distribution of molecular orientations gave a significant contribution to determine the lattice constant $a$ with respect to temperature. Figure 1 depicts the phase diagram of $CO_2$ phase I, where the red, blue, and brown lines indicate the temperatures and the pressure ranges in which the present MC simulations were performed.

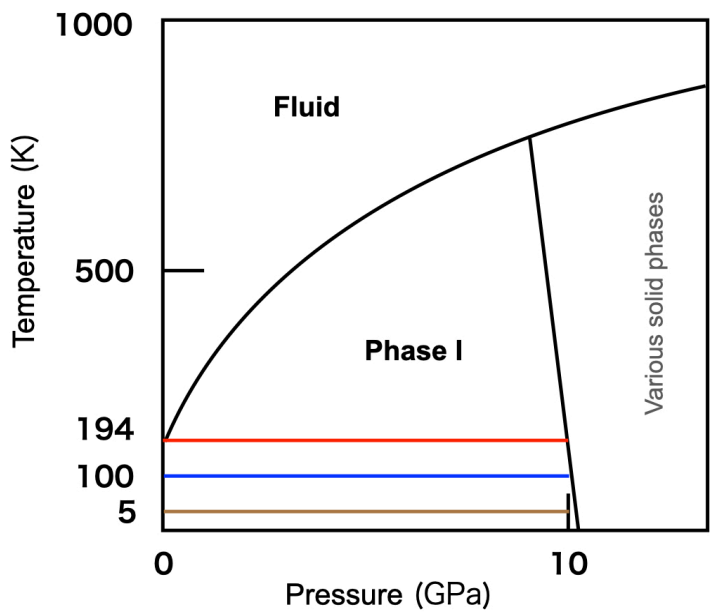

Fig. 1. Phase diagram of solid $CO_2$ phase I: red, blue, and brown lines schematically indicate the ranges of the MC simulations.

In the following, the computational procedure is described only briefly in Sec. 2. The results and discussion for $T = 5$, 100, and 194 K under $P = 1$ bar are presented in Sec. 3, which is followed by those for $T = 5$, 100, and 194 K under $P = 1$ bar and 2.5, 5.0, 7.5, and 10.0 GPa. Finally, conclusion is given in Sec. 4. In Appendices I and II, the Kihara potential and the actual computations are described, respectively.

2. Computational procedure

The basic cell is cubic and contained 2048 molecules; there are eight primitive unit cells along each edge, each primitive unit cell containing four molecules; and the edge length, $L$ ($L^3 = V$), of the basic cell was $L = 44.43$ Å at $T = 5$ K and $P = 1$ bar. The periodic boundary condition was applied to the basic cell. For intermolecular potentials, a Kihara potential of a 9-6 Lennard-Jones (LJ) type [10, see also Appendix I] was assumed to work between molecules when the molecular centers were separated by $R \leq 10$ Å. For 10 Å $< R \leq 15$ Å, only the van der Waals potential was assumed between the molecular centers (see Appendix I). The electrostatic quadrupole-quadrupole (EQQ) potential was assumed to work in the range of $R \leq 15$ Å.

3. Results and discussion

3.1 Temperature dependence of lattice constants

Figure 2 shows the temperature dependence of the lattice constants at $P = 1$ bar. The blue dots are the present results, and the blue line is only an eye guide. The lattice constants, $a$, were best fitted to the following Eq. 2:

$$a = 5.55 + 4 \times 10^{-4}\, T + 5 \times 10^{-6}\, T^2 \qquad (2)$$

The brown dots and line in Fig. 2 are the experimental lattice constant, $a_0$, shown in Eq. 1. The blue line for the calculated lattice constants $a$ appears to be steeper than the experimental line but the agreement is fairly well as seen in Eqs. 1 and 2. The $T$-dependent term of Eq. 2 gave only an insignificant contribution to $a$. The larger $T^2$ dependence of $a$ in Eq. 2 led to a large value, $a = 5.82$ Å at $T = 194$ K, that had not been obtained in the preliminary simulations based on the algorithm of the previous article [1]. It was not certain, however, that this was only due to the difference in the algorithm used.

In order to see the effects of molecular orientations on the lattice constant $a$, MC simulations were undertaken by freezing the molecular orientations as defined by the $Pa3$ structure of $CO_2$ phase I, and by allowing only the molecular centers to move. The computed results are shown by the orange dots in Fig. 2. The calculated lattice constants $a_1$ was best fitted to the following Eq. 3:

$$a_1 = 5.54 + 5 \times 10^{-4} T + 10^{-6} T^2 \tag{3}$$

The coefficient of the $T^2$ term in Eq. 3 is significantly smaller than those in Eqs. 1 and 2, and hence the fitting curve appears to be almost linear. This indicates that, as seen in Fig. 2, the steeper increase in the lattice constant $a$ over a quasi-linear increase in $a_1$ with respect to temperature is attributed to the increased random distribution of molecular orientations with temperature. In other words, the increased random distribution of molecular orientations with temperature generates additional pressure in the $CO_2$ crystal, further increasing the lattice constant. In fact, the difference of the lattice constants between $a$ and $a_1$ is larger at higher temperature.

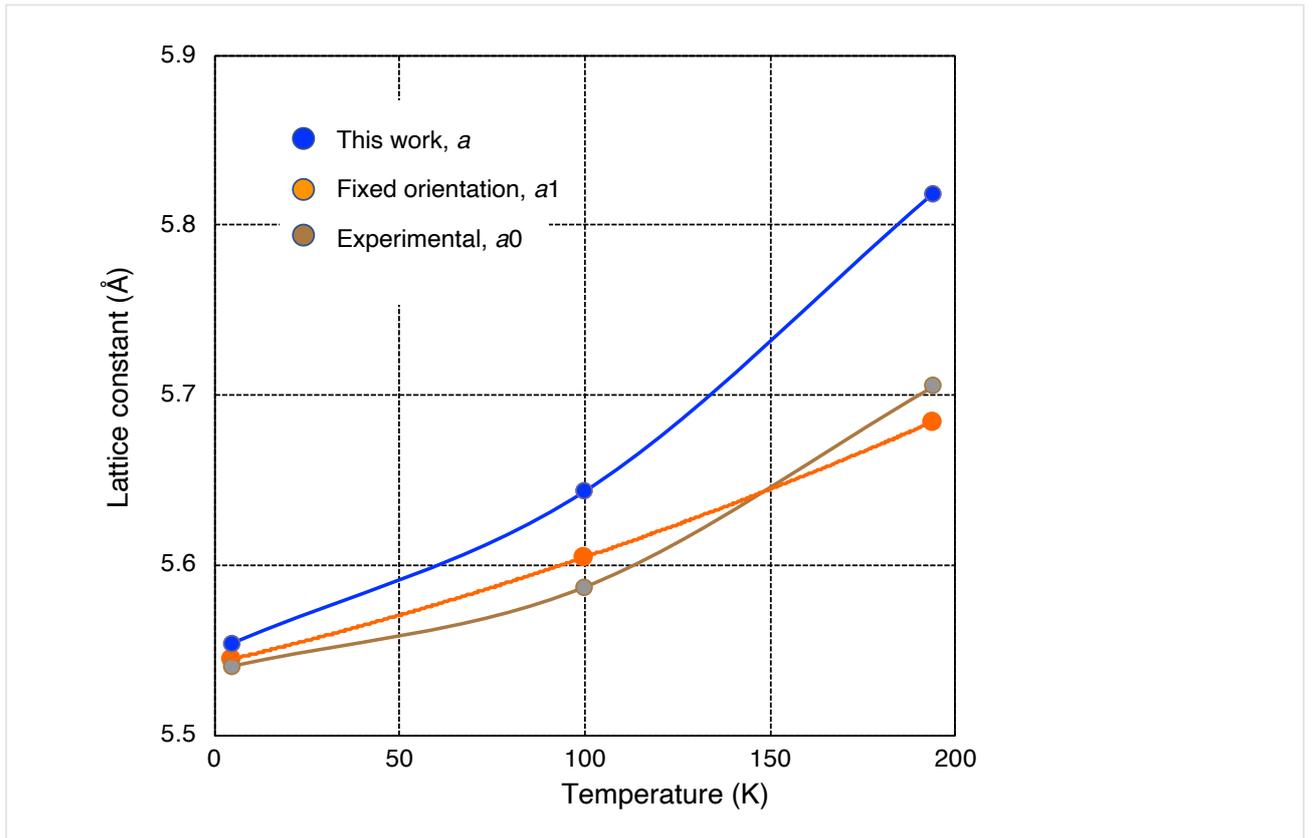

Fig. 2. Temperature dependence of lattice constants; blue dots: this work ($a$), orange dots: fixed orientation ($a_1$), and brown dots: experimental ($a_0$). The lines are only eye guide.

Figure 3 shows the molar volumes determined by the present MC simulation (blue dots), the present MC simulation with the molecular orientations fixed according to the *Pa3* structure of solid $CO_2$ (orange dots), and the experiment [6] (brown dots) that was deduced from the lattice constant $a_0$ of Eq. 1. The same discussion as above on lattice constants applies to these results: the computed results fairly agreed with experiment, and the difference between the blue dots and the orange dots are due to the random distribution of molecular orientations with temperature.

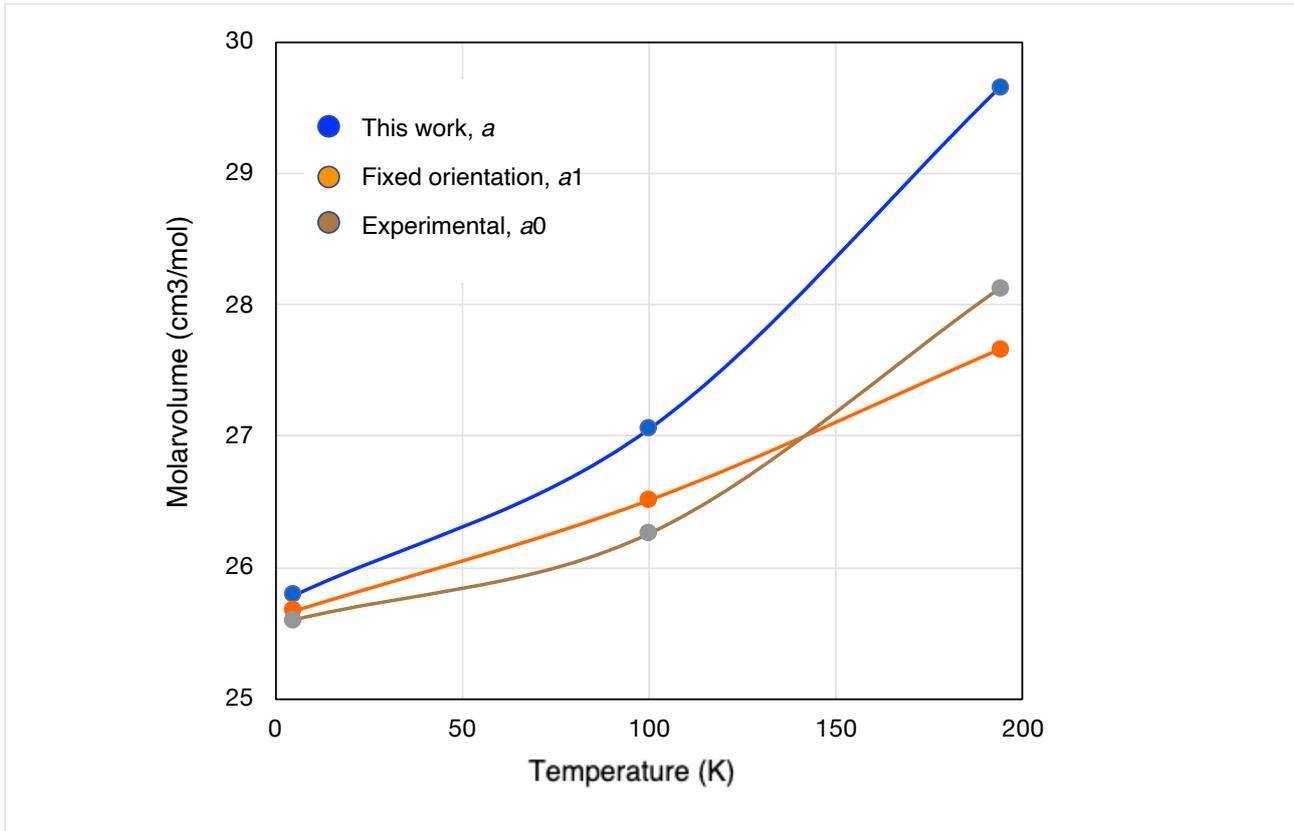

Fig. 3. Temperature dependence of molar volumes (in units of cm$^3$/mol); blue dots: this work, orange dots: fixed orientation, and brown dots: experimental. The lines are only eye guide.

3.2 Pressure dependence of lattice constants

Figure 4 depicts the pressure dependence of the lattice constants at $T = 5$ (in blue), 100 (in orange) and 194 K (in brown). The agreement with experiment [7-9] as well as *ab initio* theoretical results [11-14] was reasonably good. The difference of the lattice constants at different temperature decreased with pressure (in units of GPa) as expected.

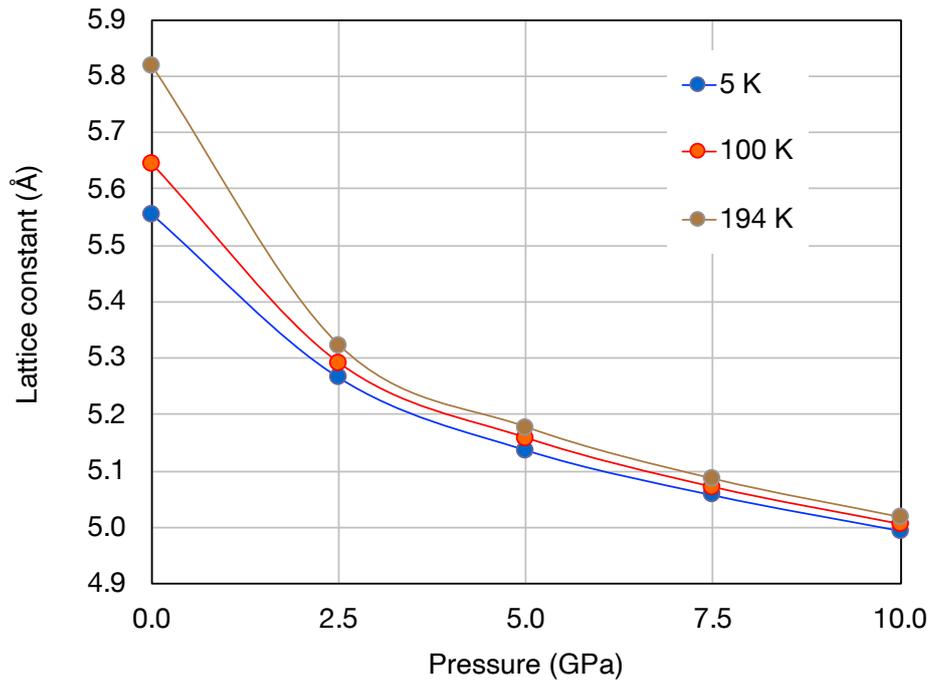

Fig. 4. Pressure dependence of lattice constants at $T$ = 5, 100, and 194 K. The lines are only eye guide.

Figure 5 shows the molar volumes (in units of cm$^3$/mol) at high pressure. Like in Fig. 4, the present results agrees reasonably well with experimental [7-9] and theoretical [11-14] results. The discussion described above also applies to this figure.

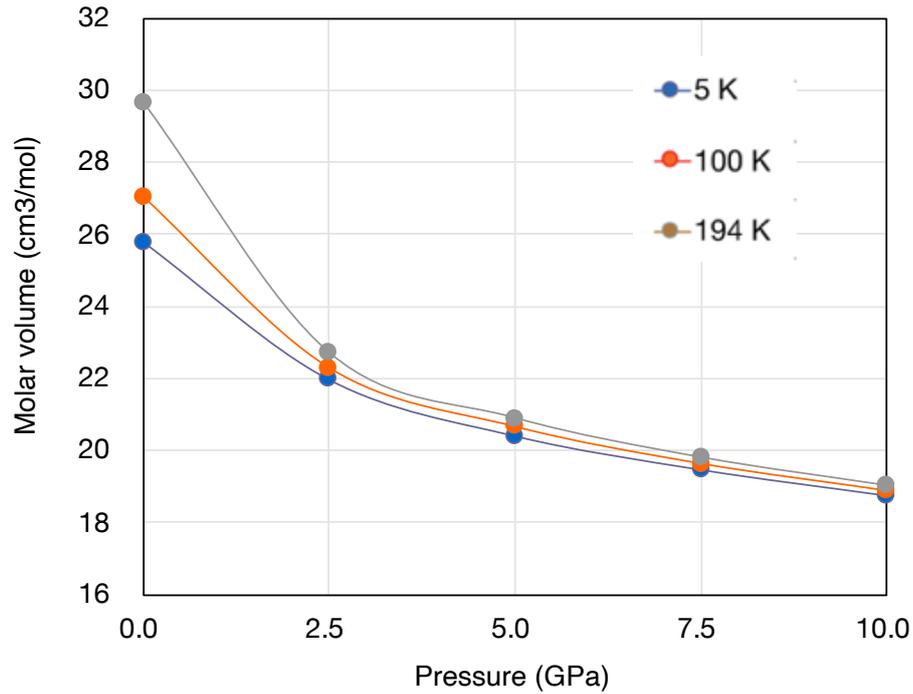

Fig. 5. Pressure dependence of molar volumes at $T$ = 5, 100, and 194 K. The lines are only eye guide.

Properties of solid $CO_2$ phase I and new solid phases at high pressure have been studied using *ab initio* density functional theory (DFT) [7-9, 11-14]. Li and his colleagues [7-9] used the second-order Møller-Plesset perturbation (MP2) method, and well reproduced the *P-V* relation, the pressure-induced phase transition between phase I and a high pressure phase III, and the Raman spectra of phase I and phase III. The research of molecular solids at high pressure using DFT is of great use especially at high pressure as the electron distribution of a molecule, and hence the intermolecular potential, is modified as molecular solids are compressed and/or crystal structures are changed. Precisely speaking, in the study of molecular solids using empirical intermolecular potentials, like in the present article, the potential parameters have to be modified with pressure so as to reproduce experimental data. In fact, in a study of the α-γ phase transition in solid $N_2$ that takes place at $P$ = 4 kbar (0.4 GPa) and $T$ = 0 K [15], both the electrostatic quadrupole moment and the parameters included in the Kihara potential (see Appendix I) had to be slightly different for the $N_2$ molecules in the α- and the γ-phases to reproduce the phase transition pressure. It was fortunate in the present article, however, that the pressure range studied was much higher than the pressure of the α-γ phase transition in solid $N_2$, hence the repulsive potential energy in the Kihara potential dominated the EQQ potential energy at high pressure, so that the computed *P-V* relation was in good agreement with experiment even without modifying the potential parameters. The present empirical intermolecular potential model allowed us to compute the temperature dependence of the molar volume as shown in Figs. 2 - 5. As a result, it was found that the random distribution of molecular orientations increased the molar volume more significantly at higher temperature. The present results indicated that the empirical potential models are still of use particularly when the random distribution of molecular orientations gives a meaningful influence on the properties of the molecular solids such as *P-V* characteristics.

4. Conclusion

The present study proved that the Kihara potential for solid $CO_2$ phase I reproduces fairly well the experimental results of *P-V* relation at the ambient and high pressures by MC simulation: the results at high pressure were consistent with those of both experiments and theoretical studies using DFT. Moreover, the present empirical potential model allowed us to compute the temperature dependence of the lattice constant at high pressure, enabling us to investigate the influence of the random distribution of molecular orientations on the *P-V* relation.

Appendix I. Kihara potential

The Kihara potential is a unique empirical model of intermolecular potential for simple molecules such as $N_2$ [15, 16] and $CO_2$ [10] as well as spherical molecules. Most often, intermolecular potential is assumed to be a sum of interatomic potentials. By contrast, the Kihara potential assumes a solid core inside a molecule, and defines the intermolecular potential as a function of the nearest core-core distance, $\rho$, as shown in Fig. A1. For the case of $CO_2$, the core is a rod with a length *l* and zero diameter.

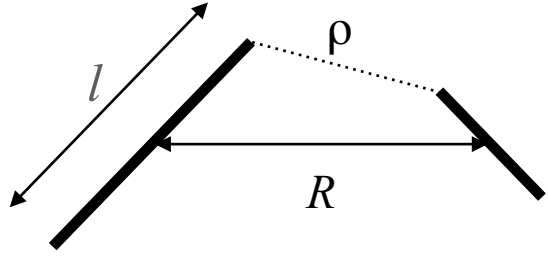

Fig. A1. Concept of Kihara potential for $CO_2$. $l$: core length, $R$: molecular center-to-center distance, and $\rho$: the shortest core-core distance.

The Kihara potential is expressed by the LJ-type potential $V_c(\rho)$,

$$V_c(\rho) = U_0 \left[ \frac{6}{n-6} \left( \frac{\rho_0}{\rho} \right)^n - \frac{n}{n-6} \left( \frac{\rho_0}{\rho} \right)^6 \right], \tag{A1}$$

where $V_c(\rho)$ expresses the 12-6 LJ potential when $n = 12$, and the 9-6 LJ potential when $n = 9$. Five sets of the parameter values for $l$, $\rho_0$, $U_0$ and $n$ have been examined in Ref. 10 in reference to the Raman spectra of solid $CO_2$ and their pressure dependence using empirical lattice constants. In the present article, $n = 9$, $U_0 = 232$ K (1 K = 1.38 x $10^{-23}$ J), $\rho_0 = 3.27$ Å, and $l = 2.21$ Å (model V of Ref. 10) were used. It should be noted that the core length $l$ used in the present simulation was smaller than the experimental O-O distance, 2.30 Å. In the actual computations, the maximum range of the Kihara potential was limited to 10 Å, and the van der Waals potential,

$$V_c(\rho) = - U_0 \left[ \frac{n}{n-6} \left( \frac{\rho_0}{R} \right)^6 \right], \tag{A2}$$

was assumed [10] up to $R = 15$ Å. The EQQ potential between $CO_2$ molecules were expressed by a point electrostatic quadrupole moment, Q = - 4.3 x $10^{-26}$ esu cm², that was placed in the center of molecule and assumed to work up to $R = 15$ Å. For the *Pa3* structure with $a = 5.54$

Å, the EQQ potential energy accounted for 43% of the energy of the cubic basic cell, a similar magnitude of the Kihara core potential energy. The EQQ potential is strongly orientation dependent, and determines the molecular orientations in the *Pa3* structure.

The Kihara potential was initially used in the research of second virial coefficient, and confirmed to be a good empirical potential for spherical and non-spherical molecules [17-27]. This justified the Kihara potential to be used for the study of molecular solids. For solid $CO_2$ phase I with a lattice constant of 5.54 Å (see Eq. 1), $\rho = 3.15$ Å between the nearest neighbor molecules, while the molecular center distance is $R = 3.92$ Å. The difference between the two numbers becomes smaller as the molecular center distance increases. To compare between the Kihara potential and the atom-atom potential, consider a $CO_2$ molecule in phase I. In the Kihara potential, the molecule interacts with the surrounding 12 nearest-neighbor molecules with 12 distances of $\rho$. This is in contrast to a $CO_2$ atom-atom potential, in which case the intermolecular potential is a sum of nine atom-atom interactions, and a $CO_2$ molecule in phase I interacts with the surrounding 12 nearest-neighbor molecules through 108 interactions. In spite of such a difference between the Kihara and the atom-atom potentials, the Kihara potential was successful to reproduce experimental results of solid $CO_2$ and $N_2$ [15, 16].

Appendix II. Numerical calculations

The parameters included in the MC simulation of the present article were (i) the volume $V$ of the basic cell that contains 2048 molecules and (ii) the molecular coordinates including the molecular center $R_i = \{x_i, y_i, z_i\}$ and the molecular orientation $\Omega_i = \{\theta_i, \varphi_i\}$ of the *i*-th molecule, where *i* runs from 1 through 2048. The coordinates relating to the molecule *i*, $\{R_i, \Omega_i\}$, will be denoted altogether as $\Psi_i = \{R_i, \Omega_i\}$. According to the MC algorithm described in the text book of Frenkel and Smit [5], a variable was randomly chosen from $\{V, \Psi_i\}$ for a MC trial, and replaced with a new value: for the volume $V$ of the basic cell, a value that was randomly chosen within the range from $-V_m$ to $+V_m$, $V_m$ being a predetermined parameter, was added to the natural logarithm of $V$: for $\Psi_i$, new values were determined using a set of predetermined parameters, $\Delta\Psi_i = \{\Delta x_i, \Delta y_i, \Delta z_i, \Delta\theta_i, \Delta\varphi_i\}$ in a similar way. The new values of $V$ and $\Psi_i$ were accepted/rejected according to the MC algorithm. The values of $V_m$ and $\Delta\Psi_i$ were determined so that the energy of the basic cell converges smoothly and rapidly to the equilibrium value. In the present computation, the values of the predetermined parameters were the same for all molecules: most often, $V_m = 0.05$, $\Delta x_i = \Delta y_i = \Delta z_i = 0.05$ Å and $\Delta \theta_i = \Delta \varphi_i = 30°$. The most serious problem encountered in the present computations was that $V$ did not change unless proper values were chosen for $V_m$ and $\Delta \Psi$: it was as if the system was trapped into a local energy minimum. Once this happened, it was difficult to find an appropriate set of parameter values to get out of the situation. Often times, the computation had been restarted from the beginning according the most common flow of the present simulation: (i) given the lattice constant $a$, calculate the energy and other parameters in the $Pa3$ structure of solid $CO_2$ phase I, (ii) randomize both the center positions and the orientations of all molecules within predetermined ranges, and (iii) undertake constant-$NPT$ MC simulation at a given pressure and a temperature. Alternatively, it was effective to restart the MC simulation from the results that ended successfully, regardless of initial temperature and pressure.

In practical works, given a pressure and a temperature, computations were continued by adjusting the values of $V_m$ and $\Delta \Psi$ until the computation ended successfully. In the last computational run, a sufficient number (on the order of $10^8$ depending on the temperature and the pressure) of MC trials was performed so that the number of acceptance in $V$ was more than 100, and then the averages of $a$ and the energy of the basic cell were calculated to designate them to the equilibrium values. The convergence of the values for $a$ and the energy of the basic cell was confirmed by calculating the standard deviations of the last 100 values, which were usually more than three orders of magnitude smaller than the average values.

The computational method used in the previous article [1] is not rigorously correct mathematically, but it seemed that the method was quite pragmatic because all molecules were subject to MC trials in every cycle. In order to reduce the CPU time and avoid the problems described above, it would be perhaps efficient to use this procedure first and then switch to the present procedure when the energy of the basic cell becomes stable, indicating that the system is close to the equilibrium. Finally, the present computations were done using GNU gfortran on ordinary desktop computers. Most often, the number of MC trials in a single run was $5 \times 10^7$ that took 200 - 300 minutes of computation time though it strongly depended on the predetermined parameter values of $V_m$ and $\Delta \Psi_i$.